\documentstyle[twocolumn,eqsecnum,aps]{revtex}
\def\gtorder{\mathrel{\raise.3ex\hbox{$>$}\mkern-14mu
             \lower0.6ex\hbox{$\sim$}}}
\def\ltorder{\mathrel{\raise.3ex\hbox{$<$}\mkern-14mu
             \lower0.6ex\hbox{$\sim$}}}
\def\msun{\hbox{$\hbox{M}_{\odot}$}}

\begin{document}
\draft
\title{Loss cone: past, present and future
}
\author{Steinn Sigurdsson}
\address{
525 Davey Laboratory \\
Department of Astronomy \& Astrophysics\\
and the Center for Gravitational Wave Physics,\\
Pennsylvania State University\\
University Park, Pa 16802\\
}
\date{\today}
\maketitle
\begin{abstract}
The capture and subsequent in--spiral of compact stellar remnants
by central massive black holes, is one of the more interesting
likely sources of gravitational radiation detectable by LISA.
The relevant stellar population includes stellar mass black holes, and possibly
intermediate mass black holes,
generally on initially eccentric orbits.
Predicted detectable rates of capture are highly uncertain, but may be high
enough that source confusion is an issue. Foreground events with relatively high
signal-to-noise ratio may provide important tests of general relativity.
I review the rate estimates in the literature, and the apparent discrepancy between
different authors' estimates, and discuss some of the relevant uncertainties and
physical processes.
The white dwarf mergers rate are uncertain by a factor of few; the neutron star
merger rate is completely uncertain and likely to be small; the black hole
merger rate is likely to be dominant for detectable mergers and is
uncertain by at least two orders of magnitude, largely due to unknown physical
conditions and processes.
The primary difference in rate estimates is due to different initial conditions
and less directly due to different estimates of key physical processes, assumed 
in different model scenarios for in-spiral and capture. 
\end{abstract}

\pacs{0430, 0480, 0490
}

\narrowtext
\section{Introduction
}
\label{sec:level1}

The characteristic frequency to which LISA \cite{da93,fo98} is sensitive 
is comparable to the orbital
frequency at the innermost stable orbit of Schwarzschild black
holes of mass $M_{BH} \sim 10^6 \msun$ and associated Schwarzschild
radius, $r_S$.
One of the more promising ``guaranteed'' sources
for LISA is the gravitational radiation from the final stages of coalescence of 
low mass ($1$--$1000\, \msun$) compact objects with low mass
massive black holes ($M_{BH} \sim 10^{6\pm 1} \, \msun$).
White dwarfs, with masses of $0.5$-$1.3 \msun$; neutron stars, with masses
of $1.4\msun$; and, stellar mass black holes, with masses from order solar masses
upwards of $100 \msun$, may all coalesce with central supermassive black holes.
The stellar population on the most tightly bound, low angular momentum orbits
in the cusp around the central supermassive black hole is depleted by prompt swallowing
of the stars ``initially'' on such orbits. The net flux of stars into the black hole
is then set by the rate at which stars (stellar remnants) can enter the ``loss cone'' \cite{fr76}
on orbits such that gravitational radiation will shrink the semi-major axis more rapidly
than other processes can increase the energy of the object on that orbit.
Stellar dynamical processes, and possibly star formation processes, determine the instantaneous and 
mean rate at which (compact) stars can enter the gravitational radiation loss cone and
evolve dynamically to coalescence with the central black hole.

Previous papers \cite{hi95,sr97,ss97,ss98,me00,fr01,fr02,iv02} considered the likely
rate for detectable signals from degenerate compact objects in the cusps
of normal galaxies, coalescing with central black holes such
as the one inferred to be present in the Milky Way.
Estimates for detectable signal rates are of the order one per
year and higher, from these sources, but the estimates are sensitive to
systematic uncertainties in the population contributing to
the signal, with very large (many orders of magnitude) cumulative formal
uncertainties in the expected signal rate. Since high rates of coalescence are
self--limiting (through depletion of low mass compact objects at
high coalescence rates; or the growth of the primary through
coalescences, to the point where the frequency of the innermost
stable orbit is too low for LISA to be sensitive to further
coalescences), the formal uncertainty in the detectable rate
is skewed to lower rates of coalescence
than the ``canonical'' estimates in the literature.
The true rate of detectable coalescences is sensitive
to the assumptions about the mass function of the primary black
hole, the number density of stars in regions where the appropriate
mass black holes are found, the mass and number of compact remnants
available for coalescence over a Hubble time in such regions, and the
large scale structure of the stellar system in which the central supermassive
black hole is embedded.

\subsection{Assumptions}

A number of similar assumptions are made by authors in estimating coalescence rates:

\begin{itemize}
\item{\bf Galactic cusps:} Galaxies tend to have ``{\it cuspy}'', non-isothermal centres.
Observations and theory suggest that real galaxies have rising density profiles
at small radii with:

$$\lim _{r\rightarrow 0} \rho \propto r^{-\gamma}$$

and $\gamma = 3/2 + p$, $p \in \{-1,+1\}$ \cite{qu95,ko95,pe72,bw76,yo80}.
Cosmological simulations suggest that a moderate cusp forms from initial condition 
cold collapse in hierarchical models \cite{nfw}, with the cusp possibly modified
by subsequent black hole growth and interaction \cite{bw76,qu95}.
Alternatively, one may assume that the cusp grows with the black hole from
an initially dense, relaxed cluster of stars, with constant density profile at small radii.

\item{\bf Relaxation:} An implicit assumption is that central SMBH are near ubiquitous, 
with $M_{BH} \sim M_{Gal-Sph} \propto \sigma_{gal}^n$ \cite{ko95,ge00,tr02}, where
$\sigma $ is the dispersion of the stellar spheroid, defined at some suitable radius
far from the black hole.
Inside the black hole effective radius, $r_h \sim GM_{BH}/\sigma^2$, the
velocity dispersion is rising like $\lim_{r \rightarrow 0} \sigma (r) \propto r^{-1/2}$,
due simply to the Keplerian potential of the central black hole.
Additional assumptions may be made about the stellar velocity anisotropy, $\beta(r)$,
and the flattening or possible triaxiality of the stellar density profile 
at small and large radii \cite{qu95,hb01,hb02}.

Note that the census of low mass supermassive black holes $M_{BH} \sim 10^5 - 10^6 \msun$
of particular interest for LISA is not available, the existence and properties of
this populations is extrapolated from the local population (Milky Way and M32),
from the higher mass SMBH observed, and from circumstantial
evidence from low luminosity active galactic nuclei (AGN), consistent with accretion onto lower mass SMBH.

\item{\bf Stellar population:} The relaxation time of the stellar population is $t_r \propto \sigma^3(r)/\rho(r)$, 
constant for $p=0$, which thus defines the critical cusp.

Most relevant galaxies are {\it not} relaxed in the centre, from observations.
However, low mass galaxies are more likely to have high stellar densities at $r_h$ and
steep cusps, and therefore may have short relaxation times inside $r_h$.
The characteristic galaxy has luminosity $L^*$. The galaxies most likely
to have relaxed cusps with central supermassive black holes in the right mass
range for LISA are sub-$L^*$ galaxies, less luminous than the characteristic
luminosity.
It is possible that the stellar population was initially relaxed as densities
were higher and high mass stars were present in galactic centers at early times.
The current relaxation
time may be longer than the initial relaxation time, and relaxation processes may have
imprinted on the initial conditions.

\item{\bf Mass function:} A general assumption that is made is that the stellar population is normal,
with a Salpeter or Scalo initial mass function. This may of course be wrong, and it
has been suggested that the initial mass function (IMF) in galactic centres 
or nuclear star forming regions
is biased to high masses (see Kroupa 2001 for review), which would substantially 
increase the estimated LISA rates. A further assumption is made about the white dwarf (WD)
mass function, as a function of the progenitor zero-age mains sequence (ZAMS) 
mass (usually assumed to be independent
of metallicity); the cut-off ZAMS mass for neutron star formation (also usually assumed
to be independent of metallicity); and the cut-off ZAMS mass and mass function of stellar
mass black holes formed (which ought not to be independent of metallicity, see Figer, this
Proceedings). Binarity may in principle be different in nuclear star clusters than
is observed in the field, either due to initial conditions, or due to  dynamical modifications
of the binary population. Since mass transfer processes in close binaries affect the
formation channels for compact objects, this will affect the inferred rate of coalescence.
In particular, binary interactions may reopen formation channels for classes of compact objects
which are not available for single stars, reinjecting a population that might naively be
assumed to be depleted.
Assuming a Salpeter IMF and some upper mass truncation, the number fractions of
different compact object populations are then of the order:
$f_{WD} \sim 0.1-0.3$, $f_{NS} \sim 10^{-3} f_{kick}$, $f_{BH} \sim 10^{-4}$.
$f_{kick}$ is the fraction of neutron stars receiving small enough a kick as not
to be displaced from the centre of galaxies containing low mass SMBH, estimating
from the globular cluster population of neutron stars, we might infer $f_{kick} \sim 0.1$.

\item{\bf Compact Remnants:} Compact remnants can then coalesce with SMBH
through gravitational radiation or direct capture. Capture into eccentric 
orbits with pericentres $\sim$ few $r_S$ which can ``grind down'' (whereby
the orbital semi-major axis shrinks gradually) through
episodic emissions of gravitational radiation, before they are upscattered by
a subsequent stellar encounter, provides the sought after periodic signal in
the final stages of coalescence (see Sigurdsson \& Rees 1997 for discussion).
%
%

\item{\bf The loss cone:} The {\it loss} {\it cone} - $\theta_{lc}(r)$ - is, by definition, the region where
the time scale for coalescence through gravitational radiation emission is shorter
than time scale to diffuse out of loss cone through random walks or large scatterings in 
angular momentum $J$ (or, rarely, energy, $E$).

The rate is sensitive to number density profile of stellar population in the
center of galaxies, $\rho \propto r^{-3/2 -p}$, to the mass function and to
any anisotropies in the stellar distribution function.

\end{itemize}

\subsection{\bf Past calculations}

A number of authors have estimated the rate of coalescences of compact objects 
into central supermassive black holes, using a number of different formalisms
and assumptions about the underlying ``initial conditions'':

\begin{itemize}

\item{\bf Formalisms:} The primary formalisms used follow Frank \& Rees (1976), 
Lightman \& Shapiro (1977, see also \cite{pe72,bw76,ms78}) and H\'enon (1973):

\begin{itemize}

\item{}Hils \& Bender (1995) used the Lightman \& Shapiro formalism
calculating the coalescence rate in M32 like galaxies (nucleated dwarf
ellipticals with high density cores, short core relaxation times 
and an evolved stellar population).  The rate calculated by Hils \& Bender
provides both a useful benchmark for other people's calculations, and
a good estimate for the range of event rates LISA may observe.

The Hils \& Bender rate for white dwarf-SMBH coalescence in M32 is $1.8 \times 10^{-8}$ events
per year.

\item{}Sigurdsson \& Rees 1997, and Sigurdsson 1997, 1998 used the Frank \& Rees formalism,
with the refinement of estimating also the large angle scattering rate, which may
contribute significantly to the total rate in the ``pinhole'' regime, as opposed
to the diffusion regime implicitly assumed by other formalisms. 
The pinhole regime assumes that change in angular momentum of individual stars is large
per unit dynamical time, compared to the width of the loss cone 
in angular momentum space; the diffusion regime applies to the opposite limiting 
case of small changes in angular momentum per star per unit crossing time.
S \& R '97 found a 
rate for M32 like galaxies consistent with Hils and Bender, with a somewhat higher
net coalescence rate if large angle scattering is included.
S \& R found that the cosmological rate is dominated by low mass nucleated spiral galaxies,
like the Milky Way, extending to lower (bulge) mass galaxies. Such galaxies are more
numerous than M32 analogues \cite{zi98,dr01}, may have higher central densities, 
lower SMBH masses (and surprisingly
higher coalescence rates) and ongoing star formation that replenishes the compact object
population.  

Including large angle scattering, S \& R found an event rate of $3 \times 10^{-8}$ per year
for white dwarf-SMBH coalescence in M32. Excluding the large angle coalescence their
rate agrees with Hils and Bender.

\item{}Miralda-Escude \& Gould (2000) used essentially the Frank \& Rees formalism
with the assumptions of Bahcall \& Wolf. They focused on low mass black hole coalescence, with the
assumption of a relaxed central populations, finding a steady coalescence rate for
Milky Way--like galaxies of $\sim 10^{-6}$ per year. In comparison, S \& R found an
``initial'' rate for LMBH-SMBH coalescence of $\sim 10^{-4}$ per year in cuspy spirals,
declining to a sustained merger rate of $\sim 10^{-6}-10^{-8}$ per year, depending on
the assumptions about the cusp structure and total black hole population.  With high initial
LMBH coalescence rates, the initial black hole population is rapidly depleted (predicting
a burst of coalescence at high--moderate redshift) and then settles to an asymptotic
lower rate as replenishment of the inner population matches the steady state depletion rate.

\item{}Ivanov (2002) used the Lightman \& Shapiro formalism, and correcting for the 
different WD mass function, gets a rate lower than but consistent with Hils \& Bender
for M32 like galaxies. Note that the statement in that paper 
that S \& R get a much lower rate
than Hils \& Bender is wrong and is directly contradicted by the S \& R paper. Some
care must be made in comparing the rates since the models are scaled to different
structural parameters, which are not independent. Also, differential and integrated
rates are of course not directly comparable. 

\item{} The current state--of--the--art models are by Freitag (2000,2002), using
a time explicit Monte Carlo realisation of the stellar cluster surrounding the central
black hole, doing a Fokker-Planck approximation evolution of the stellar dynamics,
including SMBH coalescence and growth, following the H\'enon formalism. 
The assumptions in Freitag's work are typically for a very massive and compact isothermal
cluster, with substantial structural evolution due to black hole growth and stellar evolution.
As a consequence, the rates in Freitag's models tend to be strongly time dependent, 
but they are consistent with those
of other authors, scaling Freitag's initial conditions to 
the appropriate late--time quasi--steady state
values typically approximated by other work.

\end{itemize}

All the calculations assume some $\gamma $ (or $p$), 
scale to $M_{BH}$ and one of $\sigma $, or
the number density of stars at the ``radius of influence'' of the black hole $n_*(r_h)$. 
Inspection of the published papers reveals that the dominant source of the difference
in the rates is the choice of ``initial conditions'';
principally the choice of evolved or unevolved stellar population (IMF, BH mass function),
initial density and whether the population is initially relaxed or not. 
All assume isotropy and sphericity.



\item{\bf Normal galaxies:}{\it Unrelaxed, nucleated sub-$L^*$ spirals dominate the rate.}

\item{\bf Sustainability:}Very high coalescence rates are self-limiting - they deplete the compact object population 
or grow the SMBH to larger mass, moving out of the LISA sensitivity range.
The coalescence rate declines with higher mass SMBH, primarily due to the anti-correlation
with stellar density and correlation with stellar dispersion. 

\item{\bf Aside - runaway growth:} we can check for runaway growth under these assumptions, to see whether the
growth of the loss cone with SMBH mass can trigger rapid mass growth, since adding some
mass $\delta M \ll M_{BH}$ causes the loss cone to expand. Runaway growth occurs if
the time averaged occupancy number of additional stars in the additional slice
of loss cone, $\delta \theta (\delta M)$, is greater than one.  
Trivially we find $\delta \theta_{lc} /\delta M \propto \theta_{lc}$, implying
exponential growth, but on inspection, with very low growth constant, so
effectively there is linear growth of the SMBH due to the expansion of the loss cone,
with a low growth rate.

The exception is for high $m_*/M_{BH}$ where $\delta \theta (\delta M)$ can be large
and therefore approach runaway conditions. So, in this approximation, we expect 
runaway growth for the lowest mass seed black hole only, which saturates when $M_{BH} \gg m_*$,
{\it or}, if there is a substantial population of very massive compact objects in the
central regions ($m_* \gtorder 100 \msun )$.

\end{itemize}

\subsection{\bf Predictions}

WD merger rates are uncertain by about a factor of three. The primary sources
of uncertainty are the relative number density of WDs given the observed stellar
luminosity density; and, whether the inner cusps are typically relaxed or not.

NS merger rates are bounded above to be no more than approximately few \% of the WD
rate, unless nuclear IMFs are very skewed to high masses, since the number
fraction of NS from a normal evolved stellar population is small, and the NS are
not massive enough to mass segregate efficiently in the cusp. Since they are more
massive than typical WDs, they may still contribute significantly to the observed rate.
However, if NS natal kicks are ubiquitous, then the NS population is displaced from
the inner cusp and the number density in the inner cusp is small and the rate negligible
at all times.

BH merger rates are uncertain by at least two orders of magnitude and in reality
are likely highly time dependent. Critical issues is whether the cusp is relaxed,
allowing an infusion of low mass BH from the outer cusp, what the number fraction
of low mass BH formed is, and whether it varies strongly with the metallicity of
the star forming gas, and whether star formation persists in the inner nuclear region
at late times (as seems likely given observations of the Milky Way).

\begin{itemize}

\item{\bf LISA detection:} All predictions in the literature are consistent with LISA detections.
That is, despite the range of coalescence rates, even the most conservative rate estimates
predict LISA will see multiple events during its planned operating lifetime.

\item{\bf Source confusion:} The highest rate predictions imply possible source confusion.

\item{} The uncertainties in the rate estimates are primarily assumption driven, 
dominated by the initial conditions
assumed, not the calculation method. In principle, the true initial conditions are constrainable
by observations.

\item{\bf Other processes:} Some significant physics are still not included in the models. There is some reason
to believe that the missing physical processes will not substantially affect the rate, and
will if anything tend to lead to enhanced coalescence rates, with one possible exception.
Significant tangential anisotropy of the inner stellar cusp could suppress
the coalescence rate.

Other issues of concern include:

{\it Direct capture vs spiral--in.}

Gravitational radiation detection requires orbits that spiral in, gradually shrinking the
semi-major axis over $O(10^5)$ orbits.
The capture is into orbits with  peribothrons of few $r_S$ ($e \sim 0.999 - 0.9999999$),
but not directly into event horizon, or rapidly perturbed into event horizon. Some fraction
of those orbits are scattered down, across the event horizon, by stellar encounters
near apobothron.

Estimates of direct loss fraction varies substantially ($ \sim 1/3$ or more).
There is some concern that this fraction could be very large, eliminating a large
fraction of the estimated sources. Calculations suggest that the loss fraction
is in fact modest \cite{al02,sr97,fr01}.

{\it Tidal disruption estimates.}
similar coalescence processes apply to main-sequence stars, but those stars generally
undergo tidal disruption far outside the Schwarzschild radius (except for the most
massive SMBH) and we should
observe stellar tidal disruption events \cite{ma99,am01}.
The tidal disruption rate scales approximately with the compact object 
coalescence rate (but see Miralda-Escude \& Gould 2000),  typically $r_{coll}$ for stars 
is much larger than for compact stellar remnants, so the density distribution of 
stars ``breaks'' at larger radii in the cusp.

Observationally, the inferred tidal disruption rate is consistent with maximum predicted rate
Donley et al (2002). Naively this would 
imply that the loss cone is generally full. Donley et al also find that the inferred tidal
disruption rate is significantly higher for AGNs, suggesting that dynamical processes refilling
the loss cone, or nuclear star formation, may substantially enhance the rate, contributing
strongly to the integrated cosmological rate.

\end{itemize}

\section{\bf Future issues}

A number of open issues remain to be explored in more detail.

\subsection{Brownian wandering}

The formalisms used generally assume the SMBH is fixed centrally.

In reality there can be substantial ``Brownian motion'' of the central SMBH,
particularly for the lower mass SMBH of particular interest to LISA.
In general we have some black hole velocity, $v_{BH}$; 
mean black hole displacement from the center of mass of the 
stellar population, $\langle r_{BH} \rangle$; and some time scale on which
the Brownian motion occurs, $t_{BH}$.

If the stellar population is isothermal and relaxed, then the problem is trivial,
the SMBH is in equipartition and moving as a thermally excited massive particle 
in a harmonic oscillator potential with,  $v_{iso-BH} = \sqrt{m_*/M_{BH} }\sigma $.
Clearly there are two limits:

\begin{itemize}

\item{}If $M_{BH} \approx m_*$ there is no loss cone.

\item{}If $M_{BH} \rightarrow \infty $ then loss cone is fixed.

\end{itemize}

For non-isothermal distributions, this is a hard problem. Even for
flat density, non-isothermal distributions the problem becomes non-trivial \cite{ch02}.

Clearly for a real cusp composed of a distinct population of roughly stellar mass objects,
the black hole moves. We want to estimate the mean free path, time scale for wandering and 
whether the black holes
``carry the cusp with them'' allowing an adiabatic response of the stars to the black hole
displacement.
We know that non-isothermal cusps in general have $v_{BH} = \eta v_{iso-BH}$, $\eta \gtorder 1$
(cf Chatterjee et al 2002 \cite{ch02}).

We can assume ballistic loss cone exit. ie consider the problem in the limit where the
black hole moves on some ballistic trajectory relative to the centre of mass of the stars,
changing directions and velocity amplitude only at discrete intervals, with the change
occurring rapidly compared to the duration of ballistic motion. A compact stellar remnant
is generally captured if it comes within $(10 r_S)$ of the black hole. If the the black hole
has been displaced by a comparable amount, then a ``new loss cone'' has been entered.
The time scale for this to occur is:

$$t_{lc} \approx 10 r_S/\eta v_{iso-BH} \propto M_{BH}^{3/2}$$

For a $\sim 10^6 \msun$ BH embedded in a $\sigma \approx 100$ km/sec cusp, 
we find $t_{lc} \sim 1/\eta 10^8 $ seconds,
or about 1 year. The time scale is longer if the mean-free path is shorter, since in that 
case the black hole has to random walk
out of loss cone, rather than being displaced ballistically.
%

The motion should be ballistic on time scales comparable to some fraction of the cusp
dynamical time, very roughly, $\sim 10^{2-3}$ years.

So, orbits with time scale less than 1 year are carried with the BH as it wanders.
Which are orbits inside $\sim 100 AU$ for our canonical $10^6 \msun $SMBH.

This is $\sim 10^4 r_S$, which is precisely where the dominant influx of compact
objects is coming from. 

Therefore BH wandering is significant, but possibly not dominant, with a critical
mass close to $10^6 \msun$ - more massive SMBH wander slowly and carry cusp with them,
less massive SMBH leave loss cone ballistically.

It is critical to calculate $\eta $ for non-isothermal cusps in order to estimate
a robust coalescence rate. This may require numerical simulations.

We also want the RMS displacement $\langle r_{BH} \rangle$,
as well as the
mean free path $r_{BH}$.
The black hole is effectively fixed to the centre when $r_{BH} \ll r_S$.

\subsection{Future broken symmetries}

\begin{itemize}

\item{\bf Triaxiality.}\\
If we can't move the SMBH, can we move the stars?\\
If the nucleus is triaxial, $J$ is not conserved for individual
stars on orbits - box orbits and boxlets walk through centre.
Close to the SMBH, orbits are perturbed Keplerian orbits,
precessing ellipses, the transition is inside $r_h$, at what radius the perturbed
Keplerian limit is appropriate is important.

\item{\bf Chaos:}\\ Scattering of stars by SMBH can lead to dynamical chaos where
the only integral of motion conserved for stars moving in the mean potential
of the galaxy is the energy, ignoring long time scale relaxation processes.\\
Ergodic orbits $\rightarrow$ spherical potential (cf Gerhard \& Binney 1985).\\
However, triaxial systems can be constructed with a central SMBH in which
chaos is suppressed and triaxiality persist to small radii, well inside $r_h$
(see Holley--Bockelmann et al 2002).

Regular--regular scattering, allows continued centrophilic orbits, and a
persistent flux of stars to very small radii on dynamical rather than diffusive time scales.
This is a hard problem, and in particular we don't know what real galaxies do yet, the existence
of theoretical models with suppressed chaos is not sufficient to show that real galaxies
can achieve this limit, and chaos may dominate the evolution of the centres of real galaxies
with supermassive black holes.

We need to determine flux to low $J$ from high $E$ (large $r$) and the
time scales for evolution of potential shape (cf Zhao et al 2002 \cite{zh02}).

\item{\bf Mergers:}\\ 
A significant fraction of galaxies has undergone mergers with other galaxies, including 
low mass satellite galaxies. Merger processes provide for strongly non-axisymmetric time dependent
central potentials which leads to refilling of the loss cone on dynamical time scales,
for many cusp dynamical times. Most of the integrated rate may be from post-merger galaxies
rather than quasi-static passively evolving galaxies. 
After the stellar population involved in a galaxy merger has settled into a stationary
distribution, the SMBH brought together by the collision may evolve towards 
coalescence through dynamical processes. At late stages of SMBH--SMBH coalescence
is driven by gravitational radiation, and therefore the inner stellar cusp around
each of the SMBH coming together will see a strongly time dependent potential.
This may drive significant numbers of stars into the gravitational radiation loss-cone
leading to substantial enhancement of the rate of coalescence of low mass 
objects during the process
of binary merger.

AGN activity may correlate with galaxy mergers, and coalescence may thus be
enhanced during AGN phases. This may lead to significant fraction of coalescences occurring
when the SMBH has a substantial gaseous disk, and the torquing of the inner accretion stream
during late stages of in-spiral, producing a low amplitude periodic flickering of the AGN,
which would provide a valuable electro--magnetic signature for the counterpart to the 
gravitational radiation emission.

\item{\bf Spin:}\\
Wilson \& Colbert \cite{wc95} noted that major mergers may both lead to near maximal
spin black holes, and cause abrupt changes in the spin parameter of the black holes.
Black hole spin will change the shape and extent of the loss-cone, but at a level
small compared to other uncertainties. However, the waveform of gravitational radiation
emitted during inspiral 
is sensitive to the spin of the SMBH \cite{gl02}.

Two issues arise: one is whether we can predict the fraction of near maximal Kerr
black holes from observed statistics of accreting systems and models of major mergers;
and, whether the observations of spin parameters from inspiraling compact objects
can test models of SMBH evolution. Both issues are largely open and need to be
explored further.


\item{\bf Star Formation:}\\
The largest outstanding uncertainty are the IMF, the low mass BH mass function, 
and whether star formation is ongoing in galactic nuclei, and if so, with what duty cycle.,

\item{\bf Pop III:}\\ If the Pop III IMF is peaked strongly to high mass \cite{ab02,br99}, then
at high z, intermediate mass black holes are strong contributors
to the coalescence rate and observable LISA event rate.

\item{\bf Black Hole Mass Function:}\\ We need the zero mass cut-off masses, frequency and distribution of masses for
low mass black holes as function of the IMF and metallicity (see Fryer this proceedings
and references therein). 
Knowing the binarity and initial spin of massive stars and their compact remnant
descendants would be nice too.

\item{\bf Ongoing star formation:}\\ If there is ongoing ``normal'' nuclear star formation
in inner 1-2 pc of normal spirals, with
as little as 1\% duty cycle, then compact objects are replenished and
high coalescence rate sustained. That is, if there is star formation in galactic nuclei,
and if the mass function of the stars formed is similar to that of stars formed in the
field, or possibly biased towards high mass stars, then the nuclear star formation
is important for generating a new population of compact stars in the nuclear regions. 
This is a particularly important process
if low mass black holes continue to form in the metal rich gas found
in galactic centres at late times. 

There is some observational evidence for ongoing central star formation 
in the Milky Way \cite{ec96,ge97,gh02}.
An open question is whether there is in situ formation, or off centre clusters
that then sink to the centre \cite{po94,fi98,se98}. 
We need to understand the formation channel and IMF of stars formed from metal rich 
gas deep in galactic potentials, in possibly intense radiation environments, 
near the central black holes. 

\end{itemize}

\section{Conclusions}

\begin{itemize}

\item{} Coalescence of low mass compact objects with SMBH is certain.

\item{} Rates are uncertain but consistent and optimistically high.

\item{} The outstanding model uncertainties are resolvable in principle.

\end{itemize}

The coalescence of low mass compact stellar remnants with central supermassive
black holes in galaxies is one of the major hoped for LISA sources, and detection
of such events with reasonable S/N can provide strong tests of both astrophysical
processes and fundamental physics.

LISA will hopefully test some of our models,
some model parameters will probably remain free until LISA constrains them.


\acknowledgments
This research was supported in part by NSF grant PHY-0203046
and the Center for Gravitational Wave Physics, an NSF funded Physics Frontier Center
at Penn State.

\end{document}